\documentclass[prl,aps,twocolumn,showpacs]{revtex4}
\usepackage{graphicx,bm}
\begin{document}
\title{
  Ultrahigh harmonics from laser-assisted ion-atom collisions}
\author{Manfred Lein}
\affiliation{Max Planck Institute for the Physics of Complex Systems, 
  N\"othnitzer Stra{\ss}e 38, D-01187 Dresden, Germany}
\author{Jan M. Rost}
\affiliation{Max Planck Institute for the Physics of Complex Systems, 
  N\"othnitzer Stra{\ss}e 38, D-01187 Dresden, Germany}
\date{\today}
\begin{abstract}
  We present a theoretical analysis of
  high-order harmonic generation from
  ion-atom collisions in the presence of linearly polarized 
  intense laser pulses.
  Photons with frequencies significantly higher than in standard 
  atomic high-harmonic generation are emitted. 
  These harmonics are due to two different mechanisms:
  (i) collisional electron capture and subsequent
  laser-driven transfer of an electron between projectile and target
  atom;
  (ii) reflection of a laser-driven electron from the projectile
  leading to recombination at the parent atom.
\end{abstract}
\maketitle
Over the last decades, a vast amount of work has been devoted 
to the study of \mbox{ion-atom} collisions \cite{Doerner} 
and atoms in intense laser fields \cite{Proto,Lambr}.
However, the two areas were almost entirely separated.
No experiments on ion-atom collisions in the presence of 
strong laser pulses have been carried out. The reported
experiments on laser-assisted collisions \cite{Debarre} 
involve one-photon processes and thermal collision energies.
Also, theoretical descriptions \cite{charge} have mostly 
been limited to slow collisions and/or relatively weak fields.
Recently, however, the theoretical works by 
Madsen {\em et al.} \cite{Madsen} 
and by Kirchner \cite{Kirchner} investigate fast collisions in the presence of
a strong laser field.
In Ref.~\cite{Madsen}, excitation mechanisms are discussed, 
while Ref.~\cite{Kirchner} focuses on ionization and electron capture. 
In both cases, the presence of the field leads to a significant modification of 
the collision process.
On the other hand, there has been no study on the question
how typical strong-field processes in atoms, such as high-order harmonic
generation (HHG) \cite{McPherson} and above-threshold ionization \cite{ATI}, 
are modified due to the impact of an ion projectile. In HHG, a large number 
of incoming laser photons are converted into a single high-energy photon.
HHG experiments are presently pursued with great effort \cite{Spielmann,Paul} 
since the process serves as a source  of coherent XUV radiation and 
attosecond pulses.

In the present work, we investigate HHG in laser-assisted ion-atom collision.
We focus on impact velocities such that the time-scales of nuclear and 
electronic 
motion are comparable, i.e., we have significant probabilities for collisional 
electron transfer from the target to the projectile.
For sufficiently long laser pulse durations,
the laser-driven electron effectively sees a 
large range of internuclear distances during one laser pulse.
When the laser polarization axis is parallel to the ion impact velocity, we 
show that this situation results 
in the generation of high harmonics with photon energies 
much higher than usually obtained in atomic HHG.

The classical recollision model \cite{Corkum} describes atomic HHG as a 
sequence of strong-field ionization, acceleration of 
the electron in the laser field and recombination with the core.
Within this model,
the maximum return energy of the laser-driven electron is $3.17 U_{\rm p}$ 
where
$U_{\rm p}=E_0^2/(4\omega^2)$ is the ponderomotive potential for a laser with
field amplitude $E_0$ and frequency $\omega$.
The maximum energy of the emitted photons is then equal to 
$3.17 U_{\rm p}+I_{\rm p}$ where $I_{\rm p}$ is the atomic ionization 
potential.
For ion-atom collisions, we show below how 
the interplay between collisional electron capture and laser-driven electron 
transfer between target and projectile leads to new mechanisms of HHG with 
cutoffs at significantly higher energies.

We consider collisions of protons on hydrogen atoms for proton energies of
2~keV (impact velocity $v=0.283$~a.u.). 
Due to the large impact momentum, the projectile trajectory 
is assumed to be classical and along a straight line.
Furthermore, we use a two-dimensional model where all dynamics is 
restricted to the plane that contains the target nucleus and the
projectile.
The internuclear vector is $\mathbf{R}(t) = (X,Z)=(b,vt)$ where $b$ is 
the impact parameter and $v$ is the impact velocity.
The interaction with the laser field 
\mbox{$\mathbf{E}(t) = \mathbf{E}_0(t)\sin(\omega t+\delta)$}
 is treated in the dipole 
approximation and in velocity gauge. 
The time-dependent Hamiltonian then reads (atomic units are used throughout)
\begin{equation}
  H(t) = 
  {\mathbf{p}^2\over2} + \mathbf{p}\cdot\mathbf{A}(t)
  + V(\mathbf{r}_{\rm t})
  + V(\mathbf{r}_{\rm p}),
\end{equation}
where $\mathbf{A}(t)=-\int_0^{t}\mathbf{E}(t')dt'$,
$\mathbf{r}_{\rm t}=\mathbf{r}+{\mathbf{R}(t)/2}$, and
$\mathbf{r}_{\rm p}=\mathbf{r}-{\mathbf{R}(t)/2}$.
For the electron-proton interaction $V$ we use the softcore 
potential from Ref.~\cite{Proto2}. 
For the laser field, we choose a wavelength of 800~nm and an intensity of 
1$\times$10$^{14}$~W/cm$^2$ since these parameters are readily available in
experiment. We use a trapezoidal pulse shape with a three-cycle turn on and 
turn off. The total pulse duration is 16 optical cycles, which is equal to 42.7~fs.

Initially, the electron
is in the ground state of the target atom, i.e., localized around
$\mathbf{r}_{\rm t}=0$. The time-dependent Schr\"odinger equation is 
then solved numerically by means of the split-operator 
method \cite{Feit}. The initial distance 
between target and projectile is set to 
$\mathbf{R}_0 = (b,-250{\ \rm a.u.})$
so that the closest approach occurs at mid-pulse.

The HHG spectrum is calculated according to \cite{Burnett}
\begin{equation}
  S(\omega) \sim \bigg|\int \langle \mathbf{a}(t)\rangle e^{i\omega t}dt\bigg|^2,
\end{equation}
where $\langle\mathbf{a}(t)\rangle$ is the time-dependent dipole acceleration 
vector.

\begin{figure}[t]
  \begin{center}
    \includegraphics[width=.97\columnwidth]{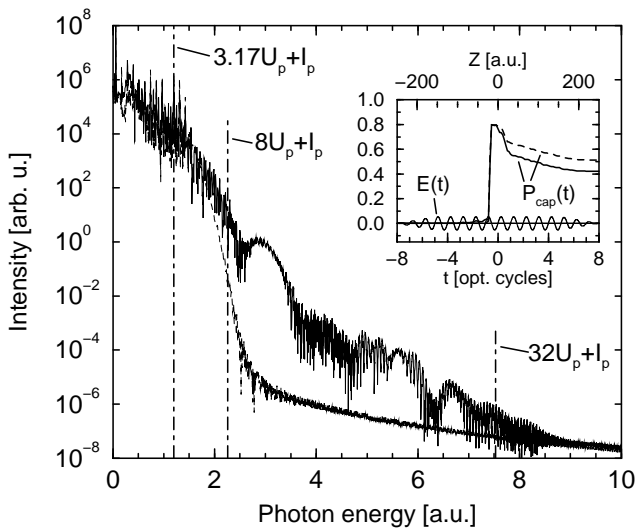}
    \caption{
      Emission spectra for impact parameter
      $b=4$~a.u. and laser phase $\delta=0$. 
      Full curve: polarization axis parallel to impact velocity. 
      Dashed curve: polarization axis perpendicular to impact velocity. 
      The vertical lines indicate the position of cutoff energies for various 
      mechanisms (see text). The inset shows the time-dependence of the 
      laser field and the capture probability for the two polarization directions.}
    \label{fig1}
  \end{center}
\end{figure}
Figure \ref{fig1} shows the harmonic spectrum obtained for a collision with
impact parameter
$b=4$~a.u. and phase $\delta=0$ of the laser. 
The two curves correspond to different directions of the laser
polarization. If the polarization axis is perpendicular to the impact velocity,
the emission spectrum has a form which is familiar from HHG in isolated atoms:
a cutoff occurs at the photon energy $3.17 U_{\rm p}+I_{\rm p}$.
Apparently, the passing projectile does not change the cutoff energy.
If the polarization axis is parallel to the impact velocity, the result is
strikingly different. 
We find an extension of HHG to frequencies reaching slightly
beyond $32 U_{\rm p}+I_{\rm p}$. 
The region between $3.17U_{\rm p}+I_{\rm p}$ and $8U_{\rm p}+I_{\rm p}$
appears like an extension of the atomic plateau with a steeper
slope.
(The significance of the values $8U_{\rm p}$ and $32 U_{\rm p}$ will be 
explained below.) Furthermore, interesting hump 
structures appear in the spectrum 
around 3~a.u., 5.6~a.u., and 6.6~a.u.

The inset of Fig.~\ref{fig1} shows the electric field $E(t)$
and the capture probability $P_{\rm cap}(t)$ which we define as the probability
that the electron is found within a square of size $40\times40$~a.u. around
the projectile. We see that the duration of the collisional capture process is 
much shorter 
than the pulse length since the ion-atom collision is essentially an attosecond
process, cf. \cite{Kirchner}. After capture at mid-pulse, 
$P_{\rm cap}$ decreases due to ionization in the strong field. 
The capture probability is smaller for the case where the laser polarization
is along the impact velocity, indicating that this geometry leads to 
more ionization during the ion-atom collision.

\begin{figure}[t]
  \begin{center}
    \includegraphics[width=\columnwidth]{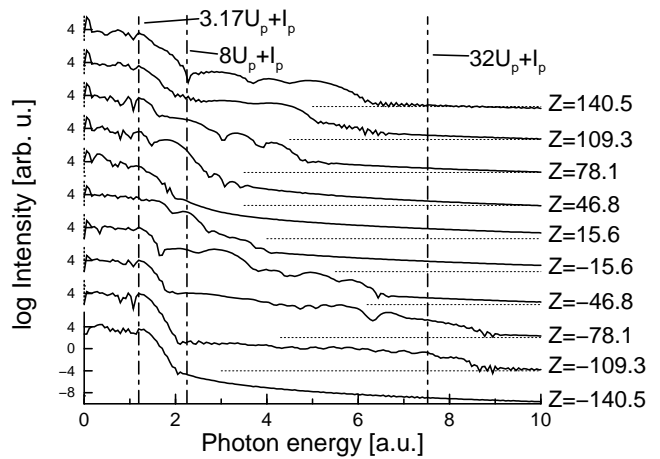}
    \caption{
      Temporal analysis of harmonic emission
      for the same parameters as in Fig.~\ref{fig1}. The
      polarization axis is parallel to the impact velocity.
      Each spectrum describes the emission during one laser cycle
      and the value $Z=vt$ refers to the middle of the respective
      time interval.
      The dotted horizontal lines indicate the respective 
      level of \mbox{$\log I = -10$}.
      The vertical lines indicate 
      the same cutoff energies as in Fig.~\ref{fig1}
      }
    \label{fig2}
  \end{center}
\end{figure}
Figure \ref{fig2} gives a temporal analysis of the harmonic emission
for the case that the laser is polarized parallel to the impact direction.
Each spectrum in the figure is obtained by Fourier transforming the dipole
acceleration over one laser cycle. 
We find that the emission below the
atomic cutoff is nearly independent of time. 
The situation is not the same for the higher
harmonics. Initially, there is no emission of ultrahigh harmonics  
($Z=-140.5$~a.u.). 
Nevertheless, harmonics up to the highest frequency are produced
already at $Z=-109.3$~a.u. and $Z=-78.1$~a.u., 
i.e. long before the actual ion-atom collision.
Around $Z=0$, the emission at the highest frequencies is weak.
Instead, harmonics up to $8 U_{\rm p}+I_p$ are generated. At later times,
emission at these energies drops and again, higher frequencies are produced.
It is evident that the spectral structures at
3~a.u., 5.6~a.u., and 6.6~a.u. in Fig.~\ref{fig1} 
mainly arise from emission around
$Z=-46.8$~a.u. and $Z=-78.1$~a.u. Since the generation of these
harmonics is limited to almost only one laser cycle, 
individual harmonics are not well
resolved in those parts of the spectrum.
The fact that ultrahigh harmonics are already generated at times when
the approaching ion is farther than 100~a.u. from the target
clearly shows that the presence of the ion influences the dynamics of the
laser-driven electrons when they are {\em far away from the target atom}.

As the next step in our analysis, 
we vary the impact parameter while keeping
all other parameters constant. 
\begin{figure}[t]
  \begin{center}
    \includegraphics[width=\columnwidth]{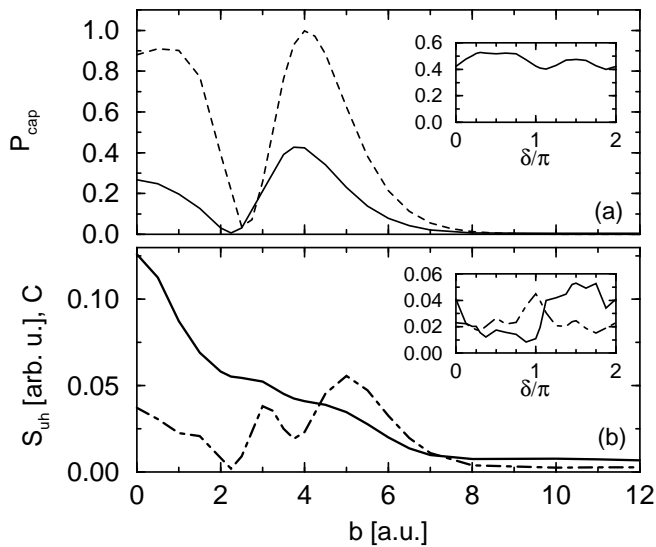}
    \caption{
      Panel (a): full curve, capture probability versus impact parameter
      for a laser field with phase $\delta=0$;
      dashed curve, field-free capture probability.
      Panel (b): 
      yield of ultrahigh harmonics (full curve) and coherence parameter 
      (dot-dashed curve) versus impact parameter.
      The polarization axis is parallel to the impact velocity.
      The insets show the same quantities as a function
      of the laser phase $\delta$ for fixed impact parameter
      $b=4$~a.u.
      }
    \label{fig3}
  \end{center}
\end{figure}
Figure \ref{fig3}(a) displays the final capture probability 
$P_{\rm cap}$ as a function of the impact parameter,
with and without the presence of the laser.
The capture probability is 
significantly reduced by strong-field ionization. The
oscillatory dependence on the impact parameter, however, is 
qualitatively unchanged.

Since we are particularly interested in 
harmonic generation well beyond the atomic cutoff, we define the
yield of ultrahigh harmonics as the integrated quantity
\begin{equation}
  S_{\rm uh} = \int_{5U_{\rm p}+I_{\rm p}}^{\infty}S(\omega)d\omega.
\end{equation}
The solid line in Fig. \ref{fig3}(b) displays the yield of ultrahigh 
harmonics as a function of the impact parameter. 
Its overall structure is a monotonic decrease. Thus, at first sight it exhibits
no correlation with the capture probability. A closer look, however, reveals
that the ``dips'' around $b=2.5$~a.u. and $b=4$~a.u. coincide with extrema
of the capture probability. For further investigation of this point we define
the coherence parameter $C$ according to $C = P_{\rm cap}P_{\rm t}$.
Here, $P_{\rm t}$ is the final probability that the electron remains bound to
the target atom, 
and is calculated analogous to $P_{\rm cap}$. Small values
of $C$ indicate either ionization or that the electron is localized at
one of the two nuclei.
For large values of $C$, the electron state is
a coherent superposition of target and projectile states.
Figure~\ref{fig3}(b) shows that the maxima of $C$ coincide with shoulders
in $S_{\rm uh}$ while the minima in $C$ coincide with the dips in 
$S_{\rm uh}$.

We conclude that the generation
of harmonics beyond the atomic cutoff originates from at least two different
mechanisms. One mechanism is unrelated to the coherence parameter
and gives rise to the monotonic decreasing background
in Fig.~\ref{fig3}(b).
A second mechanism depends on the strength of the
coherence parameter and gives rise to 
the oscillations on top of the background.

The insets in Fig.~\ref{fig3} show the dependence on 
the laser phase $\delta$ for
fixed impact parameter $b=4$~a.u. 
As we vary the phase,
we find only modest changes in the 
capture probability, but a significant alteration of $S_{\rm uh}$.
Note that phases between $\pi$ and $2\pi$ give a higher yield than
phases between $0$ and $\pi$.
No clear correlation with the coherence parameter $C$ is observed,
indicating that the first mechanism dominates.
This is consistent with the smallness of the coherence parameter 
at $b=4$~a.u.

As an explanation of our results, we propose the following two  
mechanisms of HHG. 
In mechanism (i),
one of two collision partners is ionized by the laser, the free electron is 
then
accelerated in the time-dependent laser field, and finally the electron 
recombines with the {\em other} ion. 
Mechanism (ii) also begins with the creation and acceleration of a free 
electron.
Instead of recombining at the other ion, however, the electron is elastically
reflected and is further accelerated in the laser field before it finally
recombines with the core from which it was originally ejected. 

Some years ago, mechanism (i) has
been suggested as a new mechanism of HHG in stretched molecules 
\cite{Moreno,Bandrauk,Kopold}, and a cutoff at $8U_{\rm p}+I_{\rm p}$
was derived for the special internuclear distances $R=(2n+1)\pi\alpha$, 
$n=0,1\dots$, where $\alpha=E_0/\omega^2$ is the classical quivering amplitude
of the laser-driven electron. Until now, this cutoff has not been observed in 
experiment. One reason seems to be that rather large internuclear
distances are required. Moreover, the effect occurs only for systems
where the state of the electron prior to ionization 
can be described by a single-particle orbital that is coherently 
delocalized over both nuclei. 
For example, 
this is a valid description
for the ground-state of an H$_2^+$ molecular ion, but not for a neutral
molecule at large internuclear distances. 
In an ion-atom collision, a coherent superposition is realized 
if the coherence parameter $C$ is appreciable.
This explains the connection between $C$ and $S_{\rm uh}$ in 
Fig.~\ref{fig3}(b).

Under the assumption of fixed nuclei, it is straightforward to derive the
maximum cutoff for both processes by inspection of the classical
electronic equation of motion,
$\ddot{\mathbf{r}}=-\mathbf{E}_0\sin\omega t$.
For an electron starting with zero velocity at 
$\mathbf{r}=0$ at time $t_0$, we have
\begin{equation}
  \dot{\mathbf{r}}(t) = (\mathbf{E}_0/\omega)(\cos\omega t-\cos\omega t_0).  
\end{equation}
Thus, the largest possible velocity equals $2E_0/\omega$, corresponding
to an energy of $8 U_{\rm p}$. If an electron with this energy recombines
at the other ion, it will generate a 
photon with energy $8 U_{\rm p}+I_{\rm p}$. This explains the cutoff law for 
mechanism (i). If the electron is instead 
elastically backscattered at time $t_1$, the velocity is thereafter 
given by
\begin{equation}
  \dot{\mathbf{r}}(t) = 
  (\mathbf{E}_0/\omega)(\cos\omega t + \cos\omega t_0 -2\cos\omega t_1), 
  \quad t>t_1,
\end{equation}
so that the maximum velocity is $4E_0/\omega$ corresponding to an energy
of $32 U_{\rm p}$. This maximum occurs for the case that
$t_0$, $t_1$, and $t$ are times of zero electric field. It is straightforward
to show that recollision with the 
maximum energy is possible if the internuclear distance 
satisfies the condition $R=3(2n+1)\pi\alpha$, $n=0,1,\dots$
Therefore, at these distances we have a cutoff at $32 U_{\rm p}+I_{\rm p}$.
For the present laser parameters, 
$\alpha=16.5$~a.u. and $U_{\rm p}=0.22$~a.u.
Although nonzero ion velocities may lead to a correction of the cutoff, we
conclude that the generation of ultrahigh harmonics 
(Figs.~\ref{fig1},\ref{fig2}) is 
well explained by the reflection mechanism.

\begin{figure}[t]
  \begin{center}
    \includegraphics[width=.85\columnwidth]{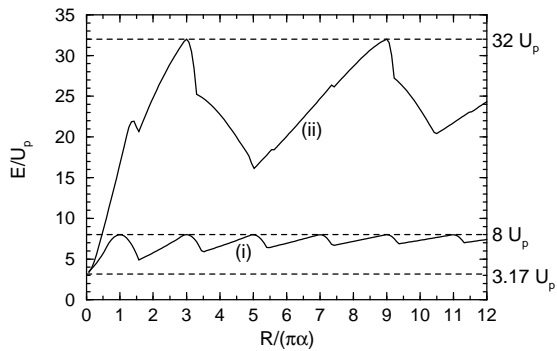}
    \caption{
      Maximum kinetic electron energy versus internuclear distance for the 
      recombination of a laser-driven electron in mechanisms (i) and (ii), see
      text (classical calculation for fixed nuclei).
      }
    \label{fig4}
  \end{center}
\end{figure}
For mechanisms (i) and (ii), 
Fig.~\ref{fig4} shows the maximum 
electron energy at recombination time as a function
of the internuclear distance. These results are obtained by numerical solution
of the classical equation of motion for fixed nuclei.
Both curves approach the atomic limit $3.17 U_{\rm p}$ for 
$R\rightarrow0$. The maximum energies are found at $32 U_{\rm p}$ and 
$8 U_{\rm p}$, respectively, and this gives rise to the cutoffs
mentioned above. Note also that already at $R=\pi\alpha$, 
the reflection mechanism 
produces electron energies above $16 U_{\rm p}$.

The phase dependence observed in Fig.~\ref{fig3} is consistent with the
reflection scenario due to the asymmetry of the ion-atom system. 
Before the ion-atom collision, reflection of electrons can occur only
for electrons that are accelerated from the target towards the projectile.
Under inversion of the field direction (phase shift by $\pi$), these
electrons are accelerated in the opposite direction where they cannot
be backscattered.

In summary, we have investigated 
ion-atom collisions in the presence of a strong laser field.
High harmonics are generated with energies much larger than found in 
atomic HHG if the laser polarization axis is parallel to the direction
of impact.  We have proposed two distinct HHG mechanisms, both involving 
laser-driven transfer of electrons between the collision partners. 
The highest harmonics are produced by reflection of electrons
from the projectile back to the target atom
at times when target and projectile are far away from each other.
A simplified classical description of the electron gives a cutoff energy 
at $32U_{\rm p}+I_{\rm p}$ for this process. In principle, the mechanisms 
of HHG that we have discussed should appear also in molecules. However,
it seems difficult to combine the required large internuclear distances
with precise control of the molecular orientation. In ion-atom
collisions, all possible distances are sampled, and the orientation control 
is straightforward.

\end{document}